\documentclass[a4paper, 12pt]{article}

\usepackage[utf8]{inputenc} 

\makeatletter
\renewcommand\footnotesize{%
   \@setfontsize\footnotesize\@xipt\@xipt
   \abovedisplayskip 10\p@ \@plus2\p@ \@minus5\p@
   \abovedisplayshortskip \z@ \@plus3\p@
   \belowdisplayshortskip 6\p@ \@plus3\p@ \@minus3\p@
   \def\@listi{\leftmargin\leftmargini
               \topsep 6\p@ \@plus2\p@ \@minus2\p@
               \parsep 3\p@ \@plus2\p@ \@minus\p@
               \itemsep \parsep}%
   \belowdisplayskip \abovedisplayskip
}
\makeatother

\usepackage[american]{babel}

\usepackage[babel]{csquotes}

\usepackage[authordate, natbib, isbn=false, backend=biber, noibid]{biblatex-chicago}
\usepackage{authblk}

\usepackage[bottom]{footmisc}

\usepackage{amsmath}
\usepackage{amsfonts}

\usepackage{caption}
\usepackage{geometry} 
\usepackage{setspace}
\usepackage{etoolbox}
\newcommand{\zerodisplayskips}{%
  \setlength{\abovedisplayskip}{0pt}%
  \setlength{\belowdisplayskip}{12pt}%
  \setlength{\abovedisplayshortskip}{0pt}%
  \setlength{\belowdisplayshortskip}{12pt}}
\appto{\normalsize}{\zerodisplayskips}
\appto{\small}{\zerodisplayskips}
\appto{\footnotesize}{\zerodisplayskips}

\usepackage{graphicx}

\usepackage{booktabs} 

\usepackage{array}

\usepackage{paralist} 

\usepackage{verbatim}

\usepackage{subfig} 

\usepackage{fancyhdr} 
\usepackage[nottoc, notlof, notlot]{tocbibind} 
\usepackage[titles, subfigure]{tocloft} 

\usepackage{comment}

\begin{document}

\thispagestyle{empty}

\begin{center}
{This is the Accepted Manuscript version of an article published by Taylor \& Francis in \textit{Statistics and Public Policy} on April 25, 2022, available with open access at: \url{https://doi.org/10.1080/2330443X.2022.2050328}. For reference, the published version should be consulted.}
\end{center}

\bigskip

\begin{center}
{\Large Policy Implications of Statistical Estimates: A General Bayesian Decision-Theoretic Model for Binary Outcomes}
\end{center}

\bigskip

\begin{center}
{\large Akisato Suzuki}

\begin{singlespace}
School of Law and Government\\
Dublin City University\\
Dublin 9, Ireland\\
\end{singlespace}
\&\\
\begin{singlespace}
School of Politics and International Relations\\
University College Dublin\\
Belfield, Dublin 4, Ireland\\
\end{singlespace}

\begin{singlespace}
akisato.suzuki@gmail.com\\
ORCID: 0000-0003-3691-0236
\end{singlespace}
\end{center}

\bigskip

\begin{center}
{\large Working paper\\ (January 27, 2022)}
\end{center}

\begin{singlespace}

\begin{center}
\textbf{Abstract}
\end{center}

\noindent
How should we evaluate the effect of a policy on the likelihood of an undesirable event, such as conflict? The significance test has three limitations. First, relying on statistical significance misses the fact that uncertainty is a continuous scale. Second, focusing on a standard point estimate overlooks the variation in plausible effect sizes. Third, the criterion of substantive significance is rarely explained or justified. A new Bayesian decision-theoretic model, ``causal binary loss function model,'' overcomes these issues. It compares the expected loss under a policy intervention with the one under no intervention. These losses are computed based on a particular range of the effect sizes of a policy, the probability mass of this effect size range, the cost of the policy, and the cost of the undesirable event the policy intends to address. The model is more applicable than common statistical decision-theoretic models using the standard loss functions or capturing costs in terms of false positives and false negatives. I exemplify the model's use through three applications and provide an R package.

\end{singlespace}


\newpage

\section{Introduction}
How should we evaluate the effect of a policy on the likelihood of an undesirable event, such as conflict? The significance test (broadly defined) examines the statistical and substantive significance of statistical estimates \autocite{Gross2015, Kruschke2018a}. It looks at whether the measure of statistical uncertainty passes a certain threshold (usually the p-value smaller than 5\% or the 95\% confidence/credible interval excluding zero), and whether the effect size is practically relevant. For example, \textcite{Quinn2007} report that peacekeeper deployment with a peace agreement, and economic development in the aftermath of civil war, have statistically significant effects and make civil war ``substantially'' less likely to recur. The study suggests these two factors are important policy options for the international community to sustain peace.

The significance test has three limitations. First, the binary criterion of statistical significance dismisses the fact that uncertainty is a continuous scale. A statistically insignificant effect can be practically significant if its uncertainty is low enough for practical reasons, if not as low as what the conventional threshold of statistical significance demands. The misunderstanding is widespread that statistical insignificance means no effect \autocite{McShane2016, McShane2017}. Reliance on statistical significance results in \textit{p}-hacking and publication bias \autocite{Simonsohn2014}; then, statistically insignificant but practically significant effects are likely to be overlooked or unreported, a lost opportunity.

Second, the effect size chosen to discuss substantive significance is usually the mean even when an interval estimate is reported. The variation in plausible effect sizes is therefore overlooked. In terms of statistical uncertainty, some effect sizes other than the mean are also probable. In terms of policy implications, while the mean is by definition computed based on all probable effect sizes and their associated probabilities, some effect sizes might be practically null values \autocite{Gross2015, Kruschke2018a}.

Third, the criterion of substantive significance is rarely explained or justified. While some suggest the importance of defining practically null values \autocite{Gross2015, Kruschke2018a}, they only suggest a context-specific utility/loss function should be developed and do not offer advice on how to do it. Researchers often claim the effect size is significant in an \textit{ad hoc} way. Referring to preset criteria such as Cohen's \autocite{Cohen1988} would not be desirable either, as it ignores contexts \autocite{Duncan2007, Harris2009}. For example, a one percentage point decrease in the likelihood of some undesirable event has different degrees of significance depending on what the event is (e.g., a national rail strike vs. civil war).

Decision theory can offer a solution to these problems by linking statistical estimates with a loss function \autocites{Berger1985, Esarey2015, Manski2019, Mudge2012}[for another solution, see][]{Suzuki2022}. For example, \textcite{McNeil1984} discuss the role of decision trees for public health policy, while \textcite{Baio2017} introduce a Bayesian cost-effective analysis for this topic. \textcite{Cecchetti2000, Warjiyo2019} present loss functions that guide macroeconomic policy. Yet, there is insufficient provision of a general decision-theoretic model whose loss function researchers can use ``off the shelf,'' to interpret the substantive significance of the statistically estimated causal effects of policy interventions for many different topics. Exceptions include \textcite{Esarey2015, Manski2019, Mudge2012, Tetenov2012}.

To advance this literature, I develop a new general Bayesian decision-theoretic model -- the causal binary loss function model \autocite[for Bayesian decision theory, see][]{Baio2017, Berger1985, Robert2007}. In summary, the model compares the expected loss under a policy intervention with the one under no intervention. These losses are computed based on a particular range of the effect sizes of a policy, the probability mass of this effect size range, the cost of the policy, and the cost of the undesirable event the policy intends to address. The causal binary loss function model is more applicable for the current purpose, than those using the standard loss functions \autocite[see][60--64]{Berger1985} or the existing general models capturing costs in terms of false positives (i.e., inferring a non-null effect when the policy actually has no effect or an adverse one) and false negatives (i.e., inferring the null effect when the policy actually has a desired effect) \autocite{Esarey2015, Manski2019, Mudge2012, Tetenov2012}. The standard loss functions are too restrictive to evaluate the effect of a policy. Modeling costs in terms of false positives and false negatives is more complicated than directly using the costs of a policy and an undesirable event.

The causal binary loss function model is better suited to evaluate the effect of a policy than the significance test, for two reasons. First, the model uses the posterior distribution of a causal effect estimated by Bayesian statistics. This enables the model to incorporate (1) a range of effect sizes other than a standard point estimate and (2) its associated probability mass as a continuous measure of uncertainty. These points address the first two limitations of the significance test.

Second, the model computes approximately up to what ratio of the cost of a policy to the cost of an undesirable event an expected loss remains smaller if the policy is implemented than if it is not. The policy is considered substantively significant up to the cost ratio where an expected loss is smaller if the policy is implemented. This point addresses the third limitation of the significance test.

It is often difficult to measure the costs of a specific policy and outcome. Yet, by definition, the cost of a policy should be smaller than the cost of an outcome, for the policy to be worth implementing. Therefore, the ratio of the former cost to the latter cost should be less than one. Thus, we can compute an expected loss given each of all possible ratios (e.g., 0.01, 0.02, …, 0.99) and simulate up to what ratio the expected loss is smaller under a policy intervention. For example, we might state: ``The policy is expected to produce a better outcome than that resulting from no policy intervention, if the ratio of the cost of the policy to the cost of the undesirable outcome is approximately up to 0.43.''

Although this may not be as informative to actual policymaking as computing an expected loss based on measured costs, it is an improvement compared to the significance test. The significance test overlooks the cost aspects when evaluating the effect of a policy intervention. Claiming that a causal factor is worth implementing as a policy, without consideration of its relative cost, would equal assuming that the policy is always cost-effective. The purpose of this article is not to use the causal binary loss function model for one specific case of policymaking, but to propose it as a better generic way than the significance test to assess the effects of policy interventions. Certainly, a model is a simplification of reality and, therefore, the causal binary loss function model should be taken as such rather than as the absolute standard for scientific reasoning and policymaking \autocite[e.g., see][]{Laber2017}. Nonetheless, as I explain in this article, the causal binary loss function model enables richer inference than the significance test (which also uses a model).

If a decision-theoretic model computes an expected loss given the estimate of a causal effect, causal identification is essential for such a model to produce a reliable result. Bayesian decision theory is not a solution to the causal identification problem. However, dismissing a decision-theoretic perspective is not a solution either; causal identification is essential for any model to discuss the effect of a policy intervention. Issues regarding causal identification are beyond the scope of this article \autocite[for useful discussion, see, for example,][]{Hernan2020, Morgan2015, Pearl2016}. The causal binary loss function model is applicable regardless of the types of causal effect estimate (population average, sample average, conditional average, individual, etc).\footnote{The model can be modified depending on the purpose of causal inference. For example, if one were interested in an average treatment effect and optimal treatment diversification for risk-averse decision makers \autocite{Manski2013}, the difference between the posterior of losses under a policy intervention and the one under no intervention could be used to compute the optimal proportion of cases to be treated.} 

I exemplify the causal binary loss function model through three applications in conflict research: one on the effect of peacekeeper deployment on the likelihood of local conflict continuation \autocite{Ruggeri2017}, another on the effect of a negative aid shock on the likelihood of armed conflict onsets \autocite{Nielsen2011}, and the other on the effect of different peace agreement designs on the likelihood of the agreement being supported by the public \autocite{Tellez2019}. The first two are observational studies, while the last one is a conjoint survey experiment. The purpose of the applications is neither to scrutinize the original studies nor to provide any substantive policy analysis or recommendation. It is only to illustrate how the model works in applied settings.

The structure of the article is as follows. First, I review the literature on decision theory and illustrate how a decision-theoretic model works in general. Then, I introduce the causal binary loss function model. Afterwards, I present the results of the application to \textcite{Ruggeri2017}; the results of the other two applications are available in the supplemental document. Finally, concluding remarks are stated with implications for future research. To facilitate the use of the causal binary loss function model, I provide an R package and a step-by-step vignette (see the section ``Supplemental Materials'').\footnote{All statistical analyses for this article were done on RStudio \autocite{RStudioTeam2020} running R version 4.1.2 \autocite{RCoreTeam2021}. The data visualization was done by the ggplot2 package \autocite{Wickham2016}.}

\section{Decision Theory}
Decision theory is formal modeling of decision-making. It can be categorized into descriptive vs. normative \autocite[chap.1]{Rapoport1998}, and theoretical vs. statistical \autocite[e.g.,][]{Berger1985, Gilboa2009}. Descriptive decision theory explains how people make decisions; normative decision theory examines how people should make decisions. To model uncertainty, theoretical decision theory makes distributional assumptions without incorporated empirical analysis, whereas statistical decision theory statistically incorporates empirics.\footnote{The theoretical literature reserves the term ``uncertainty'' for the circumstance where there is no knowledge of the state of the world, while using the term ``risk'' for the circumstance where there is partial knowledge \autocite[50]{Rapoport1998}. Risk analysis often uses ``risk'' to capture both probability and severity of outcomes \autocite[2]{Walpole2020}. This article follows the statistical terminology, where decision under uncertainty includes decision with partial knowledge.} The causal binary loss function model is a Bayesian normative statistical decision theory. An example of descriptive statistical decision theory is \textcite{Signorino1999}, who incorporates the implications of game theory into statistical modeling. Examples of Frequentist normative statistical decision theory are \textcite{Manski2019, Mudge2012, Tetenov2012}, which I discuss later.

Decision problems are framed either as a loss or as a gain. In a loss function, a loss is a positive value and the goal is to minimize it, while in a utility function, a gain is a positive value and the goal is to maximize it. The difference lies only in the framing, and it is possible to form a loss function and a utility function in a mathematically equivalent way.\footnote{Prospect theory suggests it changes decision-making whether we frame an issue at stake as a gain or as a loss even when the issue is mathematically equivalent \autocite{Kahneman1979}. This is not a problem in the causal binary loss function model, because what matters is not how much expected loss a decision produces, but which decision produces a smaller expected loss.} This article uses the loss framing.

The basic form of a loss function is:

\begin{equation}\label{genLossFun}
l=L(\theta,a),
\end{equation}

\noindent
where $l$ is a loss caused by action $a$, $L(\cdot)$ denotes a loss function, and $\theta$ is a parameter (or a set of parameters). Typically, $\theta$ is unknown and estimated as a posterior distribution computed by Bayesian statistics \autocite{Gelman2013BDA, Gill2015}. If an action $a$ minimizes the loss, it is considered as the optimal decision. For example, if a posterior distribution over the chance of rain next week implies Saturday has the highest probability of sunny weather, an outdoor event should be organized for that Saturday to minimize the (expected) loss (here, disappointment caused by bad weather) and not for any other day. However, because it is only an estimate with some probability (though the largest probability among all days), it is still possible that Saturday will be rainy.

Standard loss functions include the absolute error loss, $L(\theta,a)=|\theta-a|$; the squared error loss, $L(\theta,a)=(\theta-a)^2$; and the 0-1 loss, $L(\theta,a)=0$ if $\theta=a$ and $1$ if $\theta\neq a$ \autocite[60--64]{Baath2015, Berger1985}. In other words, as an action deviates from a parameter value, the penalty of the absolute error increases linearly (e.g., a deviation of 2 results in a loss of 2, a deviation of 4 results in a loss of 4, and so on), while the penalty of the squared error loss increases quadratically (e.g., a deviation of 2 results in a loss of 4, a deviation of 4 results in a loss of 16, and so on). The 0-1 loss is an extreme case: no penalty if a decision is correct and full penalty if not, such as cutting a correct cable to defuse a bomb \autocite{Baath2015}. These standard loss functions are usually used to evaluate the performance of statistical estimators. The quadratic loss function has also been utilized to model decision-making for macroeconomic policies, because of its tractability \autocite{Horowitz1987, Mayer2003}. An example given by the literature is (with slight differences across authors): $L=w_{1}(v - v^{*})^{2} + w_{2}(z - z^{*})^{2}$, where $v$ is output, $v^{*}$ is desired output, $z$ is an inflation rate, $z^{*}$ is a desired inflation rate, and $w_{1}$ and $w_{2}$ are weights \autocites[51]{Cecchetti2000}{Mayer2003}[162]{Warjiyo2019}.

However, none of these standard loss functions is flexible enough to interpret the implications of the statistically estimated causal effects of policy interventions. As for the squared error and absolute error losses, it is unclear what it means to minimize the difference between $a$, a decision to do the intervention, and $\theta$, the effect size of a policy intervention. The 0-1 loss is too simplistic; the costs should take more values than the two integers in policymaking contexts.

Loss functions for causal effects usually model costs in terms of false positives and false negatives. \textcite{Manski2019, Tetenov2012} offer a Frequentist approach to determine the treatment allocation that minimizes the maximum regret. For example, when there are two treatment options -- the status quo and an innovation, regret is the mean loss in terms of the population, given a certain portion of individuals is assigned an inferior treatment \autocite[302]{Manski2019}. An allocation is said to minimize the maximum regret if it produces the loss smaller than any other allocations in all possible states of the unknown world \autocite[123--124]{Manski2013}.

\textcites{Esarey2015, Mudge2012} propose Bayesian and Frequentist models respectively where the loss function suggests whether a policy should be implemented, given the statistical estimates of its effect size and uncertainty, a decision maker's desired effect size, and the (expected) costs to the decision maker of a false positive and a false negative. For example, \textcites{Esarey2015} apply the model to the effect of democracy on the likelihood of civil war. The practical implication of the model is that if the presence of democracy produces greater utility given all the aforementioned factors, democracy should be promoted to prevent civil war.

The causal binary loss function model is different from these models, in that it directly uses the costs of a policy intervention and an undesirable event. It is not necessarily straightforward to map the costs of a policy intervention and an undesirable event onto the (expected) costs of a false positive and a false negative, as shown in the supplemental document. The causal binary loss function model is not a substitute for the existing statistical decision-theoretic models, as the concept of false positives and false negatives is also important. This model is another addition to researchers' toolkit.

\section{Causal Binary Loss Function Model}

\subsection{Loss function}
Let $p(\theta | D)$ denote the posterior distribution of the effect sizes, where $\theta$ is a parameter for effect sizes and $D$ denotes data.\footnote{There is usually the implicit assumption that a posterior is also conditional on the specifications of a statistical model such as a prior choice, causal identification strategy, a functional form, etc.} Also, let $C_p$ be the cost of implementing a policy and $C_e$ be the cost of an undesirable event. $C_e$ is realized only if the undesirable event takes place. $C_{e}{\text{logit}}^{-1}(\pi)$ is the cost of the event times the baseline likelihood of it taking place (i.e., without the policy being implemented), where ${\text{logit}}^{-1}(\cdot)$ is the inverse logit function mapping log odds onto probability, and $\pi$ is the baseline log odds. To avoid confusion, I use ``likelihood'' to refer to the probability of an event and ``probability'' to refer to the probability that $\theta$ takes a certain range of values.

The causal binary loss function is:

\begin{equation}\label{myLossBinary}
\mathbb{E}[l]=C_p I(i) + C_e \text{logit}^{-1}(\pi + p \theta_{int} I(i) + q \theta_{unint} I(i)),
\end{equation}

\noindent
where $\mathbb{E}[l]$ denotes an expected loss; $\theta_{int}$ is the mean value of the range of the \textit{intended} effect sizes of a causal factor, i.e., the effect sizes each of which reduces the likelihood of an undesirable event by a certain amount (thus, $\theta_{int} < 0$ by definition); $p$ is the probability mass of this range under the entire posterior $p(\theta | D)$; $\theta_{unint}$ is the mean value of the range of the \textit{unintended} effect sizes of the causal factor, i.e., the effect sizes each of which increases the likelihood of an undesirable event by a certain amount or by a zero amount (thus, $\theta_{unint} \geq 0$ by definition)\footnote{If the policy had the null effect, it would still be undesirable because of the cost of the policy being wasted.}; $q$ is the probability mass of this range under $p(\theta | D)$; $I(i)$ is an indicator function taking a value of one if the causal factor is implemented as a policy and a value of zero if not. How to specify $p\theta_{int}$ and $q\theta_{unint}$ is explained in greater detail in the next subsection. While $p\theta_{int}$ and $q\theta_{unint}$ are derived from $p(\theta | D)$, for simplicity I omit the conditional statement $|D$.

If a policy is not implemented ($I(i)=0$), the expected loss is the cost of the event weighted by the baseline likelihood, ${\text{logit}}^{-1}(\pi)$, for example, the cost of civil war weighted by its baseline likelihood. If a policy is implemented ($I(i)=1$), the expected loss is the cost of implementing the policy plus the cost of the event weighted by the revised likelihood, $\text{logit}^{-1}(\pi + p \theta_{int} + q \theta_{unint})$. For example, if $p\theta_{int} = -0.1$ and $q\theta_{unint} = 0.02$, the revised likelihood of the undesirable event is $\text{logit}^{-1}(\pi + (-0.1) + 0.02)$.

Both $p\theta_{int}$ and $q\theta_{unint}$ are necessary \textit{ex ante} in the loss function, because if the posterior has both negative and positive values, it means the causal factor could decrease or increase the likelihood of the undesirable event. In the special case where the posterior of $\theta$ has only negative values, $q=0$ and $q\theta_{unint}I(i)$ disappears from the loss function. Similarly, in the special case where the posterior of $\theta$ has only positive values, $p=0$ and $p\theta_{int}I(i)$ disappears from the loss function. It might be asked why the loss function does not use the mean of the whole posterior rather than dividing it into the negative and positive values. In fact, the mean is a special case of specifying $p\theta_{int}$ and $q\theta_{unint}$, as I explain in the following subsection.

\subsection{How to specify $p\theta_{int}$ and $q\theta_{unint}$}
As for $\theta_{int}$, we use the mean over the range $-\infty < \theta \le \theta_{m.d.}$: $\mathbb{E}[\theta | D, \theta \le \theta_{m.d.}]$, where $\theta_{m.d.}$ is some minimum \textit{desired} effect size by which the likelihood of an undesirable event is reduced (therefore, $\theta_{m.d.}<0$ by definition). $p$ is the probability mass of this range in terms of $p(\theta|D)$. Formally: $p\theta_{int} 
=  \int_{-\infty}^{\theta_{m.d.}} p(\theta | D) d\theta \int_{-\infty}^{\theta_{m.d.}} \theta p(\theta | D, \theta \le \theta_{m.d.} ) d\theta$.

As for $\theta_{unint}$, we use the mean over the range $\theta_{m.u.} \le \theta < +\infty$: $\mathbb{E}[\theta | D, \theta_{m.u.} \le \theta]$, where $\theta_{m.u.}$ is some minimum \textit{undesired} effect size by which the likelihood of an undesirable event is increased or not changed (therefore, $\theta_{m.u.} \geq 0$ by definition). $q$ is the probability mass of this range in terms of $p(\theta|D)$. Formally: $q\theta_{unint} =  \int_{\theta_{m.u.}}^{\infty} p(\theta | D) d\theta  \int_{\theta_{m.u.}}^{\infty} \theta p(\theta | D, \theta_{m.u.} \le \theta) d\theta$.

If $\theta_{m.d.} < -\epsilon$ (where $\epsilon$ is an infinitesimal value and $\theta_{m.d.} \leq -\epsilon \equiv \theta_{m.d.} < 0$) and/or $\theta_{m.u.} > 0$, the non-null values between $\theta_{m.d.}$ and $\theta_{m.u.}$ are the region of practically null values \autocite{Gross2015, Kruschke2018a}. The mean of the entire posterior is a special case, where there is no region of practically null values, i.e., $\theta_{m.d.} = -\epsilon$ and $\theta_{m.u.} = 0$, so that $p\theta_{int} + q\theta_{unint} = \int_{-\infty}^{-\epsilon} p(\theta | D) d\theta \int_{-\infty}^{-\epsilon} \theta p(\theta | D, \theta \le -\epsilon ) d\theta +  \int_{0}^{\infty} p(\theta | D) d\theta  \int_{0}^{\infty} \theta p(\theta | D, 0 \le \theta) d\theta = \int \theta p(\theta | D) d\theta$.

We do not preset the values of $\theta_{m.d.}$ and $\theta_{m.u.}$, unless we are applying the model to a particular policymaking context and know the values of $\theta_{m.d.}$ and $\theta_{m.u.}$. Otherwise, the preset values would be as arbitrary as the statistical significance of $p<5\%$. This is why the causal binary loss function has $p\theta_{int}$ and $q\theta_{unint}$ separately, rather than the mean of the posterior only, which would preset $\theta_{m.d.}=-\epsilon$ and $\theta_{m.u.}=0$. We can vary $\theta_{m.d.}$ and $\theta_{m.u.}$ to see how the results change with respect to the comparison between the expected loss under a policy intervention and the one under no intervention.

\begin{figure}[t]
  \includegraphics[scale=0.125]{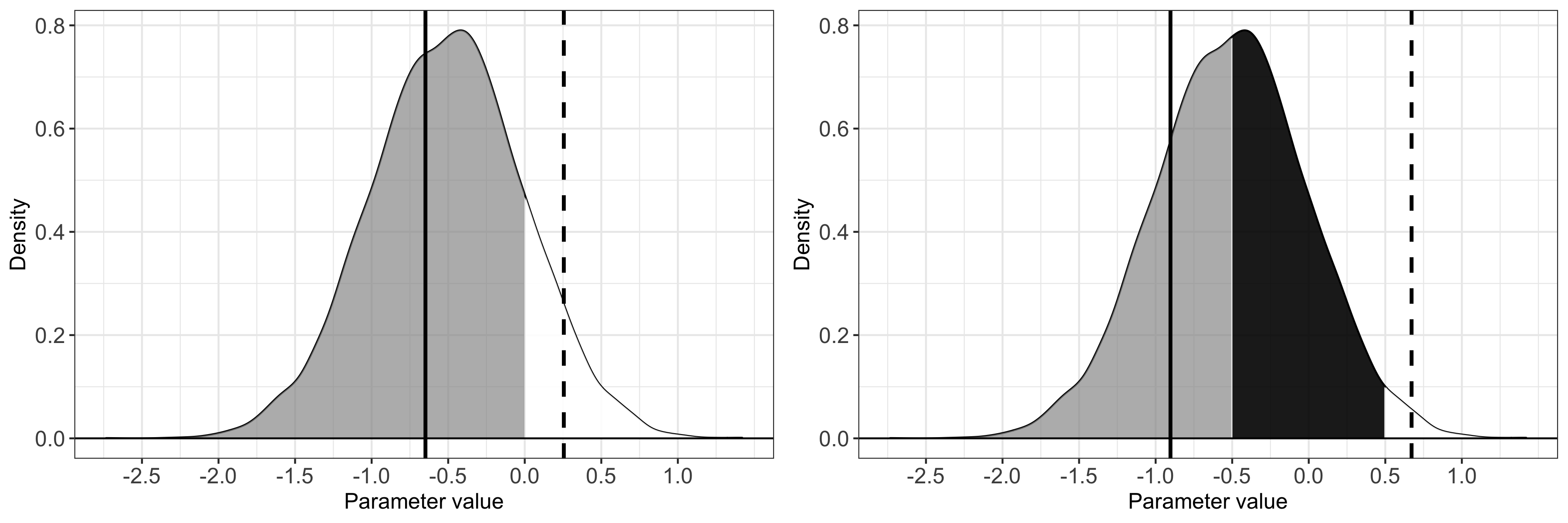}
  \centering
  \caption{Graphical example of $p\theta_{int}$ and $q\theta_{unint}$. In the left panel, the gray-shaded area is where the parameter value $<0$; the white-shaded area is where the parameter value $\geq0$. In the right panel, the gray-shaded area is where the parameter value $\le-0.5$; the black-shaded area is the region of the presumed practically null values; the white-shaded area is where the parameter value $\geq0.5$. The solid vertical line is the mean within the gray-shaded area; the dashed vertical line is the mean within the white-shaded area.}
  \label{exTheta}
\end{figure}

Figure \ref{exTheta} exemplifies what $p\theta_{int}$ and $q\theta_{unint}$ capture. The density plot is made of 10,000 draws from a normal distribution with the mean of $-0.5$ and the standard deviation of $0.5$. Let us assume the distribution is posterior samples. In the left panel, I set $\theta_{m.d.}=-\epsilon$ and $\theta_{m.u.}=0$ for exposition. For $p\theta_{int}$, the mean over the range of the effect size being smaller than zero is indicated by the solid vertical line, and its probability mass by the gray-shaded area. Formally, $p\theta_{int} =  \int_{-\infty}^{-\epsilon} p(\theta | D) d\theta \int_{-\infty}^{-\epsilon} \theta p(\theta | D, \theta \le -\epsilon) d\theta \approx -0.548$. For $q\theta_{unint}$, the mean over the range of the effect size being equal to or greater than zero is indicated by the dashed vertical line, and its probability mass by the white-shaded area. Formally, $q\theta_{unint} = \int_{0}^{+\infty} p(\theta | D) d\theta \int_{0}^{+\infty} \theta p(\theta | D, 0 \le \theta) d\theta \approx 0.04$.

In the right panel, I set $\theta_{m.d.}=-0.5$ and $\theta_{m.u.}=0.5$, again for exposition. For $p\theta_{int}$, the mean over the range of the effect size being equal to or smaller than $-0.5$ is indicated by the solid vertical line, and its probability mass by the gray-shaded area. Formally, $p\theta_{int} =  \int_{-\infty}^{-0.5} p(\theta | D) d\theta \int_{-\infty}^{-0.5} \theta p(\theta | D, \theta \le -0.5) d\theta  \approx -0.456$. For $q\theta_{unint}$, the mean over the range of the effect size being equal to or greater than 0.5 is indicated by the dashed vertical line, and its probability mass by the white-shaded area. Formally, $q\theta_{unint} = \int_{0.5}^{+\infty} p(\theta | D) d\theta \int_{0.5}^{+\infty} \theta p(\theta | D, 0.5 \le \theta) d\theta \approx 0.015$. The black-shaded area is the region of the presumed practically null values.

\subsection{How to define $C_p$ and $C_e$}
As costs in policymaking contexts are often multidimensional and unobservable, so are the cost of a policy intervention, $C_p$, and the cost of an undesirable outcome, $C_e$. For example, the cost of a policy may be different for a policymaker and for the general public; the cost may be not only financial but also political, social, and psychological. Nonetheless, even when we do not have the actual values of the costs, ignoring them as in the significance test is not ideal. Claiming that a causal factor is worth implementing as a policy, without consideration of its relative cost, would equal assuming that the policy is always cost-effective. Such an assumption may not always be true. The causal binary loss function model allows for a thought experiment: What if the ratio of $C_p$ to $C_e$ were 0.1, 0.2, 0.3, and so on? 

We can compute expected losses given the ratio of $C_p$ to $C_e$, for example, from 0.01 up to 0.99, by holding $C_p$ at 1 while varying the value of $C_e$ from 1.01 to 100 and assuming that these costs are measured on the same scale.\footnote{$C_p < 0$ is possible especially when $C_p$ is defined in a narrow, financial sense (e.g., a new policy uses less money than the status-quo policy). This article does not focus on such cases, as it usually costs more to address problems by a policy intervention than doing nothing.} Then, we can make an argument such as: ``The policy is expected to produce a better outcome than that resulting from no policy intervention, if the ratio of the cost of the policy to the cost of the undesirable outcome is approximately up to 0.43.''

Varying $C_e$ rather than $C_p$ may often allow for more informative interpretation, if it is harder to measure the cost of an undesirable event (such as war) than the cost of a policy intervention (such as sending peacekeepers), as in the case of a rare but extreme event \autocite{Taleb2007}. It would be unnecessary to consider the case where the ratio of the two costs is one, because such a case would make no difference between a policy intervention and no such intervention, even if the policy reduced the likelihood of an undesirable event to zero. Thus, the meaningful ratio is up to $1-\epsilon$.

There are caveats in defining or interpreting $C_p$ and $C_e$. The value of $C_p$ does not, and should not, include the benefit a specific policy would bring about; the benefit is captured by another term in the loss function, $p\theta_{int}$. The value of $C_e$ is not, and should not be, the one weighted by the likelihood of an undesirable event; the weighted value is $C_e{\text{logit}}^{-1}(\pi+p\theta_{int}I(i)+q\theta_{unint}I(i))$.

\subsection{Statistical setup}
There are four points to consider in setting up a statistical model to estimate $\pi$, $p\theta_{int}$, and $q\theta_{unint}$. First, $\pi$, $\theta_{int}$, and $\theta_{unint}$ should be measured on the log odds (ratio) scale, because they are inside the inverse logit function. Bayesian logistic regression estimates the posteriors of log odds ratio coefficients, which can directly be used for the loss function.\footnote{The loss function could be modified to accommodate other Bayesian models for binary outcomes. For example, if a Bayesian linear probability model were used, the inverse logit function could be replaced with the identity function.} When the effect size is interpreted, the log odds ratio might be converted to the odds ratio (by exponentiating it), or to the difference on the probability scale (by computing the difference between the predicted likelihood under a policy intervention and the one under no intervention), for ease of interpretation.

Second, the value of 1 in the binary outcome variable must mean something undesirable, so that a reduction in the baseline likelihood, ${\text{logit}}^{-1}(\pi)$, reduces $C_e$. For example, if the outcome of interest is the termination of civil war, the dependent variable should be coded 1 for no termination and 0 for termination, so that the baseline likelihood of civil war going on is reduced from $C_e{\text{logit}}^{-1}(\pi)$ to $C_e{\text{logit}}^{-1}(\pi+p\theta_{int}+q\theta_{unint})$, provided that $p\theta_{int}+q\theta_{unint}<0$, of course.

Third, a unit change in the causal variable of interest must mean a policy intervention that intends to \textit{reduce} the likelihood of an undesirable event. For example, in \textcite{Nielsen2011}, the outcome of interest is the onset of armed conflict, and the causal variable of interest is the binary indicator of a large sudden drop in foreign aid that is theoretically expected to \textit{increase} the likelihood of the onset. Then, a policy intervention to reduce this likelihood is to keep the flow of foreign aid stable (e.g., despite a budget constraint due to a financial crisis). If the binary causal variable is coded such that it takes a value of 1 for a large sudden drop in foreign aid and a value of 0 for its absence as done in \textcite{Nielsen2011}, $p\theta_{int}$ and $q\theta_{unint}$ must be derived from the posterior of the log odds ratio coefficient of the variable \textit{times} $-1$, a \textit{negative} unit change of the variable (from 1 to 0); accordingly, the baseline log odds, $\pi$, must capture the case where the causal variable takes a value of 1. If the variable is recoded to take a value of 0 for the presence of a large sudden drop in foreign aid and a value of 1 for its absence, $p\theta_{int}$, and $q\theta_{unint}$ can be derived directly from the posterior of the log odds ratio coefficient (because it equals being multiplied by $1$, a positive unit change of the variable from 0 to 1). The application of the causal binary loss function model to \textcite{Nielsen2011} in the supplemental document exemplifies the latter approach. The same logic applies to a continuous causal variable. $\theta$ should be defined as the log odds ratio coefficient times the size of the unit change of interest (e.g., a one million US dollar increase in foreign aid) that captures a policy intervention intended to reduce the likelihood of an undesirable event; $\pi$ should be defined according to what the baseline value of the continuous causal variable is.

Finally, if there is more than one explanatory variable in a statistical model, we must decide the log odds ratio coefficient of which variable to use to specify $p\theta_{int}$ and $q\theta_{unint}$, while leaving those of the other variables absorbed to $\pi$. It is the same logic as when we estimate the fitted values of an outcome variable: we manipulate the value of the treatment variable of interest while leaving the remaining covariates at some fixed values. If we estimated conditional average treatment effects or individual treatment effects and were interested in whether a policy intervention produces a smaller expected loss for a specific case, the covariates could be held at the values that represent that case. If we estimated the sample average treatment effect and were interested in whether a policy intervention on average produces a smaller expected loss, we could compute an inverse logit value for every observation of the data used for the statistical analysis, and take the average of these inverse logit values. Formally, $\frac{1}{n}\sum_{j=1}^{n}{\text{logit}}^{-1}(\pi_{j}+p\theta_{int}I(i)+q\theta_{unint}I(i))$, where $n$ is the total number of observations used for statistical analysis, $j$ indexes each observation, and $\pi_{j}$ here becomes the baseline log odds plus the dot product of the vector of the covariates for $j$ and that of their log odds ratio coefficients. It follows that the sample average treatment effect is what makes the difference between the expected loss under the policy intervention and the one under no intervention. All applications in this article take this latter approach.

\subsection{Summary and caveats}
The causal binary loss function model computes approximately up to what ratio of the cost of a policy intervention, $C_p$, to the cost of an undesirable event, $C_e$, an expected loss remains smaller when the policy intervention is done than when it is not. It utilizes a Bayesian statistical model that estimates the posterior distribution of the effect sizes of a policy. The model can vary the minimum desired and undesired effect sizes of a policy intervention ($\theta_{m.d.}$ and $\theta_{m.u.}$) to define a practically relevant range of the effect size. The continuous nature of the uncertainty of a policy's effect is captured by $p$ and $q$. $p$ is the probability of the policy realizing at least the minimum desired effect size, i.e., the probability mass of the effect size range $-\infty < \theta \le \theta_{m.d.}$ under the entire posterior $p(\theta | D)$. $q$ is the probability of the policy realizing at least the minimum undesired effect size, i.e., the probability mass of the effect size range $\theta_{m.u.} \le \theta < +\infty$ under the entire posterior.

The output of the causal binary loss function model is only an approximation for two theoretical reasons. First, the model relies on the results of a statistical model, and a statistical model itself is an approximation. Second, since a Bayesian statistical model usually estimates parameters by a Markov-chain Monte Carlo (MCMC) method, the stochastic nature of the method may well make exact results differ slightly, depending on computational settings and environments. These points affect the exactness of up to what ratio of the cost of a policy intervention to the cost of an undesirable outcome an expected loss remains smaller when the policy intervention is done than when it is not. This is why I say \textit{approximately}. The implication is that the more similar the expected losses are between when the policy is implemented and when it is not, the more sensitive the conclusion on which decision is better than the other is to the statistical model and the stochastic nature of an MCMC method. Finally, the practical reason to say ``approximately'' is that there are usually many decimal points in the results and, for communication purposes, they need to be rounded to fewer digits.

\section{Application}
\textcite{Ruggeri2017} examine the effect of peacekeeper deployment in a particular locality on the likelihood of local conflict. One of their findings is that, on average, peacekeeper deployment reduces the likelihood of local conflict continuation in its locality. I apply the causal binary loss function model to this analysis. The purpose is neither to scrutinize the original study nor to provide any substantive policy analysis or recommendation. The application is only for illustrative purposes; I take the original empirical model and causal identification strategy for granted.

\subsection{Statistical model}
The research design of the original study is as follows \autocite[169--173]{Ruggeri2017}. The unit of analysis is grid cells (approximately $55km \times 55km$) per year where violent conflict was observed in a previous year, from 1989 to 2006 in sub-Saharan Africa. The dependent variable is binary, coded 1 if a grid cell observed the continuation of local violent conflict in a year; 0 otherwise. The treatment variable is the presence of peacekeepers in a grid cell in a year, coded 1 if they were present and 0 otherwise. Thus, the policy decision is whether to deploy peacekeepers in a locality where there is ongoing violent conflict.

The original study uses weighted logistic regression adjusting for cell-specific pre-treatment covariates, on data matched on these covariates by coarsened exact matching \autocite{Iacus2012}. These covariates are the average traveling time to the nearest major city, the distance from the capital, the distance from the international borders, the infant mortality rate, the population, the average roughness of terrains, and the average rain precipitation. For greater detail, see \textcite[170-172]{Ruggeri2017}. The weights are those generated by the matching algorithm to account for different numbers of control and treated units across strata \autocite[5]{Iacus2012}.

I use the following Bayesian logistic regression model with the same set of variables as in Model 2 in the original study:

\begin{equation}\label{logitRep}
y_{j,t} \sim Bernoulli({\text{logit}}^{-1}(\beta_0+\beta_1d_{j,t}+\mathbf{x}_{j,t}\boldsymbol{\gamma})),
\end{equation}

\noindent
where the subscript $j,t$ indicates every grid cell-year observation used; $y_{j,t}$ is the dependent variable; $\beta_0$ is the constant; $\beta_1$ is the log odds ratio coefficient for $d_{j,t}$, the treatment variable; and $\mathbf{x}_{j,t}\boldsymbol{\gamma}$ is the dot product of the vector of the covariates $\mathbf{x}_{j,t}$ and that of the corresponding log odds ratio coefficients $\boldsymbol{\gamma}$.\footnote{The covariates are those already mentioned plus the cubic polynomials of conflict duration \autocite{Carter2010a}.} This setup means the effect of peacekeeper deployment is the one averaged within the data.

The posteriors are estimated by the MCMC method via Stan \autocite{StanDevelopmentTeam2019b}, implemented via the rstanarm package version 2.21.1 \autocite{Goodrich2020a}. I use four chains, each of which has 20,000 iterations; the first 1,000 iterations are discarded. I utilize the weights function in the rstanarm package to incorporate the weights generated by the matching algorithm, so that the Bayesian model becomes as close to the original Frequentist model as possible.

I use a weakly informative prior of $Normal(0,\text{log}(10))$ for the log odds ratio coefficients of all explanatory variables. The scales of these priors are then automatically adjusted based on the measurement scale of each variable by the rstanarm package \autocite{Goodrich2020a}. I use a weakly informative prior of $Normal(0,10)$ for the constant; the mean of zero makes sense, as the rstanarm package automatically centers all predictors during the estimation process -- the returned results are on the original scales \autocite{Goodrich2020a}. The weakly informative priors lead to more data-driven but also regularized Bayesian estimation. I use these priors here for reference purposes only.

\subsection{Applying the loss function}
As the binary treatment indicator is coded such that a value of 1 means peacekeeper deployment (a policy intervention to reduce the likelihood of local conflict continuation), $\beta_1$ can be directly plugged into the causal binary loss function. It will be denoted as $\theta^{p.k.}$ in the context of the loss function (where $p.k.$ stands for peacekeepers).

Let the cost of deploying peacekeepers be $C_{p.k.}$ and the cost of local conflict continuation be $C_{l.c.c.}$. Applying the causal binary loss function, we have:

\begin{equation}\label{myLossRep}
\mathbb{E}[l] = C_{p.k.}I(i) + C_{l.c.c.} \frac{1}{n} \sum_{j=1}^{n_{j}} \sum_{t=1}^{n_{t}(j)} {\text{logit}}^{-1} (\pi_{j,t} + p\theta_{int}^{p.k.}I(i) + q\theta_{unint}^{p.k.}I(i)),
\end{equation}

\noindent
where $I(i)$ is an indicator function taking a value of 1 if peacekeepers are deployed and a value of 0 if not; $n$ is the total number of observations used in the regression model; $n_{j}$ is the number of cross-section units (grid cells); $n_{t}(j)$ is the number of time units (years) for cross-section unit $j$. $\pi_{j,t}$ is the posterior mean of the constant $\beta_0$ plus the dot product of the vector of the covariates $\mathbf{x}_{j,t}$ and that of the posterior means of their corresponding log odds ratio coefficients $\boldsymbol{\gamma}$. $\theta_{int}^{p.k.}$ is the mean of the range of the intended effect sizes of peacekeepers, i.e., the effect sizes each of which reduces the likelihood of local conflict continuation by a certain amount; $p$ is the probability mass of this range under the entire posterior $p(\theta^{p.k.} | D)$. $\theta_{unint}^{p.k.}$ is the mean of the range of the unintended effect sizes of peacekeepers, i.e., the effect sizes each of which increases the likelihood of local conflict continuation by a certain amount or by a zero amount; $q$ is the probability mass of this range under $p(\theta^{p.k.} | D)$. The mean of the inverse logit values across all observations is taken, so that the average treatment effect on the probability scale is what makes the difference between the expected loss under peacekeeper deployment and the one under no deployment.

\subsection{Results}
The $\hat{R}$ was approximately 1.00 for each parameter, suggesting the Bayesian regression did not fail to converge. The effective sample size exceeded at least 30,000 for each parameter. According to the significance test, the 95\% credible interval of the log odds ratio would lead us to conclude that the effect of peacekeeper deployment is statistically significant, as the 95\% credible interval excludes zero (see Table \ref{mci}). Yet, from the perspective of the causal binary loss function, there is no imperative to use the threshold of the 95\% probability.

\begin{table}[h]
\centering
\begin{tabular}{lll}
\hline
Mean & Lower bound & Upper bound \\ 
\hline
$-1.53$ & $-2.05$ & $-1.02$\\ 
\hline
\end{tabular}
\caption{Mean and 95\% credible interval of the log odds ratio of peacekeeper deployment. $N=704$, $\hat{R}\approx1.00$ for all parameters.}
\label{mci}
\end{table}

\begin{figure}[t]
  \includegraphics[scale=0.2]{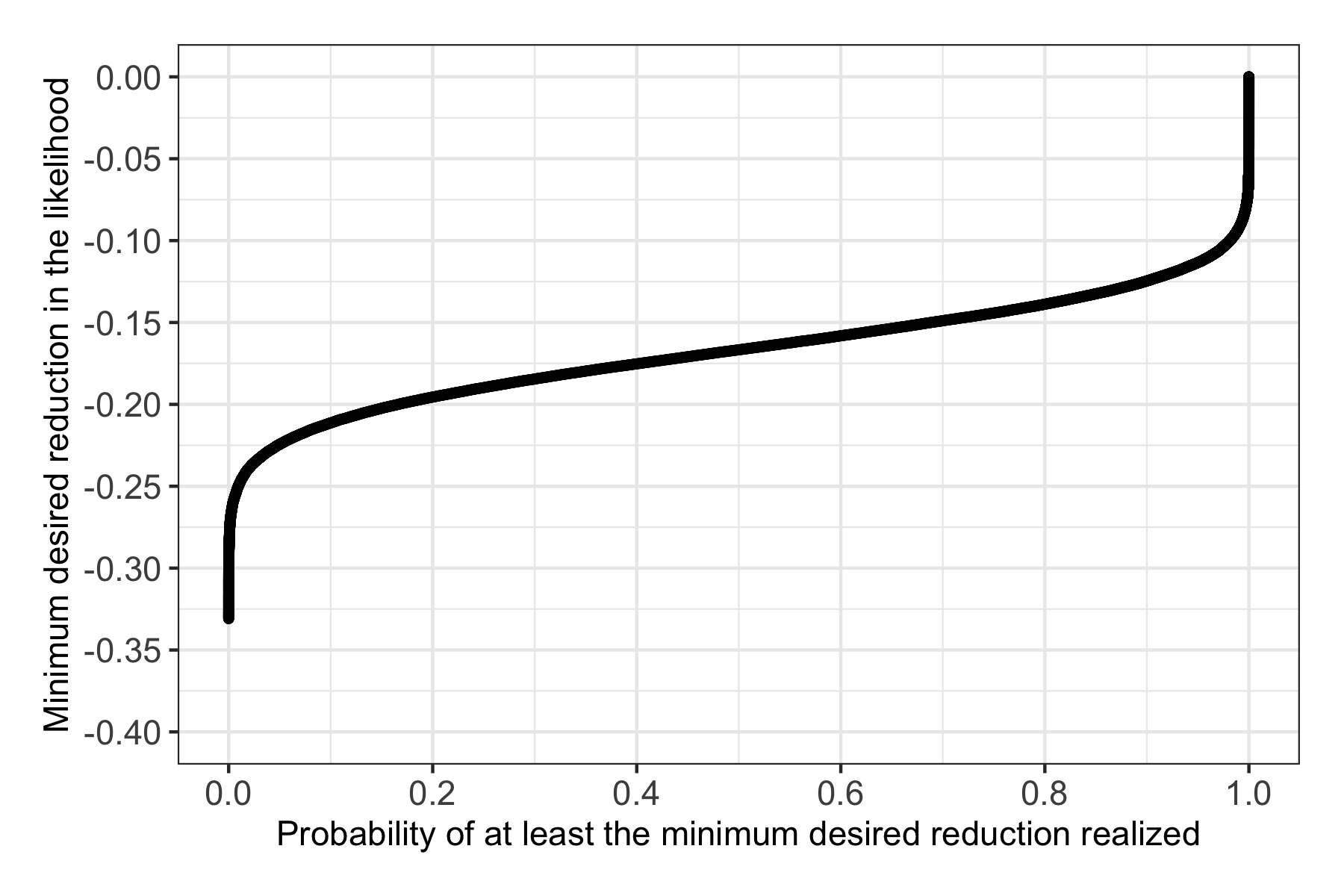}
  \centering
  \caption{Minimum desired reduction in the likelihood of local conflict continuation and the probability that peacekeeper deployment realizes such an effect. Baseline likelihood of local conflict continuation: 85\%.}
  \label{probs}
\end{figure}

Before computing the expected losses, let us focus on the relationship between the effect size range of interest and its probability mass. In Figure \ref{probs}, the y-axis shows different minimum desired effect sizes on the probability scale. The x-axis represents the probabilities that peacekeeper deployment realizes such effects. For example, if one wants to expect that peacekeeper development reduces the likelihood of local conflict continuation at least by 10 percentage points, the probability of peacekeeper deployment realizing that is approximately 98\%. If one wants to expect that peacekeeper development reduces the likelihood of local conflict continuation at least by 20 percentage points, the probability of peacekeeper deployment realizing that is approximately 17\%. In other words, the probability that peacekeeper deployment is expected to have a desirable effect depends on how much reduction is desired.

Now, let us compute the expected losses. First, as for $p\theta_{int}^{p.k.}$, I change the value of $\theta_{int}^{p.k.}$ by assuming that either a 10 percentage point or 20 percentage point reduction in the likelihood of local conflict continuation is the minimum desired effect size. These effect sizes correspond approximately to $\text{log}(0.3717)$ and $\text{log}(0.1677)$ on the log odds scale in the model. Thus, in the loss function, I define $\theta_{int}^{p.k.}$ either as $\mathbb{E}[\theta^{p.k.} | D, \theta^{p.k.} < \text{log}(0.3717)]$ or as $\mathbb{E}[\theta^{p.k.} | D, \theta^{p.k.} < \text{log}(0.1677)]$. $p$, the probability mass of the range of the desired effect sizes under $p(\theta^{p.k.}|D)$, changes accordingly. The choice of these minimum desired effect sizes is arbitrary, as the purpose here is to illustrate how different minimum desired effect sizes change up to what cost ratio an expected loss remains smaller under a policy intervention. Second, as for $q\theta_{unint}^{p.k.}$, I define $\theta_{unint}^{p.k.}$ by assuming that no reduction in the likelihood of local conflict continuation is the minimum undesired effect size. In other words, $\theta_{unint}^{p.k.} = \mathbb{E}[\theta^{p.k.} | D, 0 \le \theta^{p.k.}]$. It follows that $q$ is the probability mass of this effect size range in terms of $p(\theta^{p.k.}|D)$. Third, I change the ratio of $C_{p.k.}$ to $C_{l.c.c.}$ from 0.01 to 0.25, holding $C_{p.k.}$ at 1 while varying the value of $C_{l.c.c.}$ from 4 to 100. This range of the ratio is large enough to show the crossover point where the expected loss under the policy intervention becomes greater than the one under no intervention, and helps clearer visualization.

\begin{figure}[t]
  \includegraphics[scale=0.175]{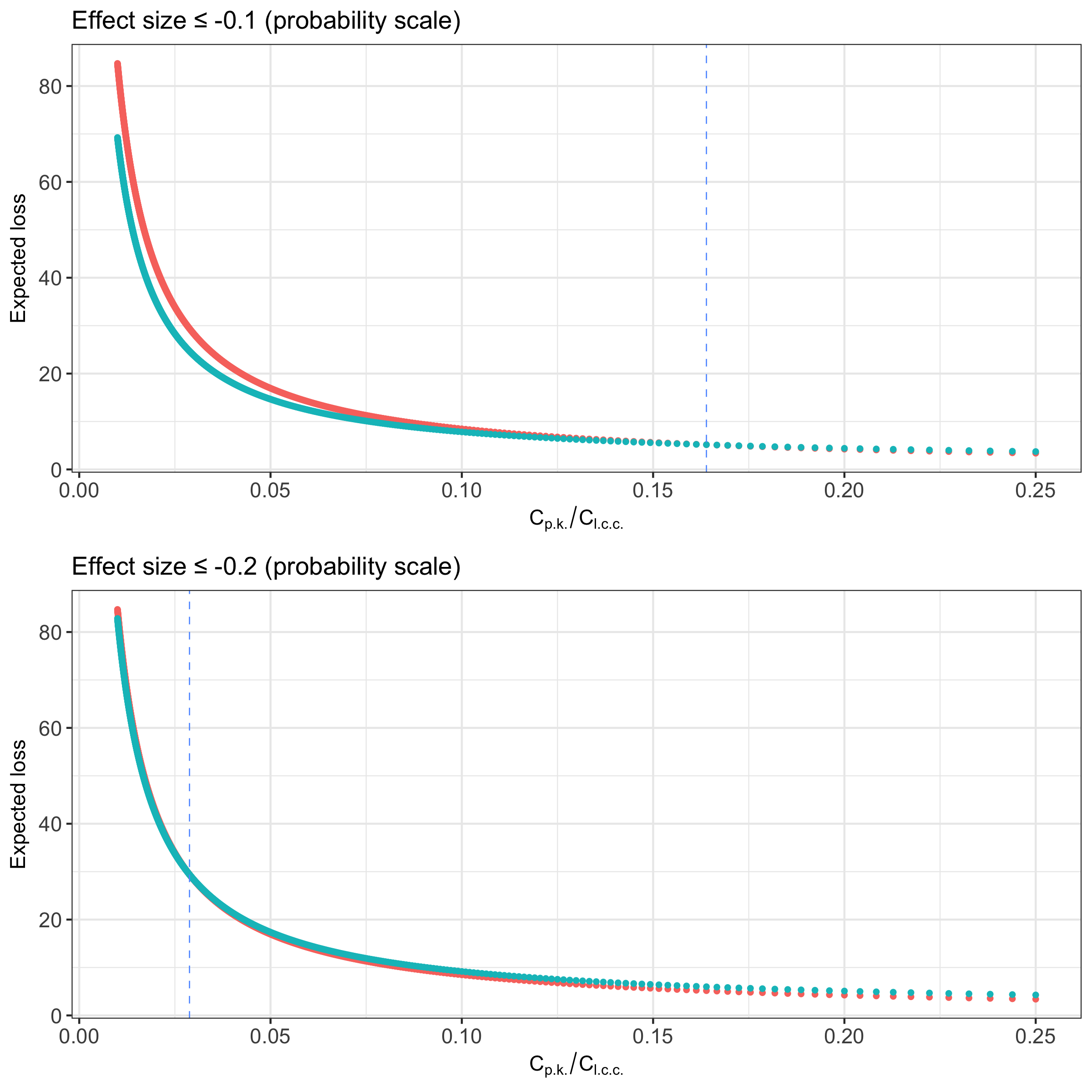}
  \centering
  \caption{Expected losses over different cost ratios. The green dots are the expected losses when peacekeepers are deployed; the red dots are the expected losses when they are not. The vertical blue dashed line indicates the cost ratio up to which the expected loss is smaller when peacekeepers are deployed than when they are not.}
  \label{losses}
\end{figure}

The results are displayed in Figure \ref{losses}. The x-axis is the ratio of $C_{p.k.}$ to $C_{l.c.c.}$. The y-axis is the expected loss. The green dots are the expected losses given $I(i)=1$, i.e., when peacekeepers are presumed to be deployed; the red dots are those given $I(i)=0$, i.e., when peacekeepers are presumed not to be deployed.

In the top panel of Figure \ref{losses}, where the 10 percentage point reduction is the minimum desired effect size, the expected losses are smaller when peacekeepers are deployed than when they are not, approximately up to the cost ratio of 0.1639 (indicated by the vertical blue dashed line). If the 20 percentage point reduction is the minimum desired effect size, as in the bottom panel of the figure, the crossover point is at a smaller value of the cost ratio, approximately at the ratio of 0.0288 (again, indicated by the vertical blue dashed line).

\begin{figure}[t]
  \includegraphics[scale=0.175]{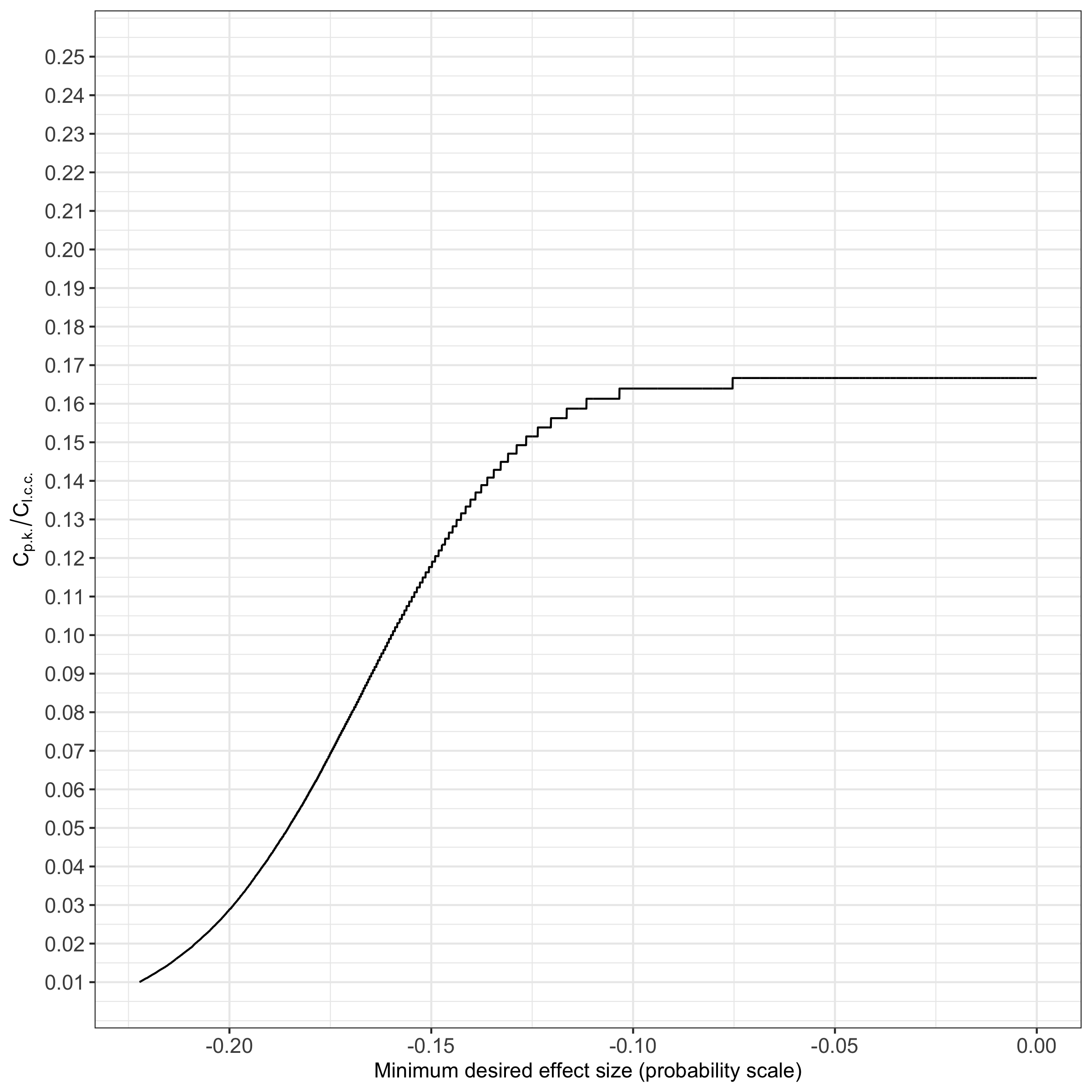}
  \centering
  \caption{Relationship between the cost ratio up to which the expected loss is smaller when peacekeepers are deployed than when they are not, and the plausible minimum desired effect sizes.}
  \label{largcrs}
\end{figure}

We can generalize these results, by computing the cost ratio up to which the expected loss is smaller when peacekeepers are deployed than when they are not, for every plausible value of the minimum desired effect sizes (by ``plausible,'' I mean the values available from the posterior samples of the causal parameter). Figure \ref{largcrs} visualizes this relationship.

To sum up, whether deploying peacekeepers produces a smaller expected loss depends on (1) the minimum desired size of reduction in the likelihood of local conflict continuation, (2) the probability that deploying peacekeepers has at least this minimum desired effect size, and (3) the relative value of $C_{p.k.}$ to $C_{l.c.c.}$. Using the causal binary loss function, we can see approximately up to what ratio of $C_{p.k.}$ to $C_{l.c.c.}$ peacekeeper deployment produces a smaller expected loss given different minimum desired effect sizes. This is richer inference over the practical implications of the effect of a policy intervention, than the significance test.

While the above discussion is about the general implications of the effect of a policy intervention, one could also consider its implications for a specific policymaking context if information about relevant actors' preference were available. Here, I discuss two points to be considered. First, a minimum desired effect size may be endogenous to the ratio of the cost of a policy to the cost of an undesirable outcome. In the current example, if the cost of deploying peacekeepers were small, policymakers might think a 10 percentage point reduction is enough, whereas if the cost of deploying peacekeepers were large, they might think a 20 percentage point reduction is necessary to justify the high cost. The process of defining the effect size of interest given the costs is outside the statistical model \autocite{French2018}. Thus, the endogeneity between the effect size of interest and the costs does not affect the statistical estimation.

Second, the costs may well differ across actors. In the current example, some countries may have better capacity to send peacekeepers efficiently than others. The cost of deploying peacekeepers is then lower to the former than to the latter. Similarly, policymakers in some countries may be more concerned with violent conflict in a particular locality than those in other countries, for example, neighboring countries vs. distant countries. The cost of local conflict continuation is then perceived to be greater by the former than by the latter.

Whether one refers to a specific policymaking context or not, the main conclusion from the causal binary loss function model holds. Whether a policy (deploying peacekeepers in the current example) produces a smaller expected loss depends on the minimum desired level of reduction in the likelihood of an undesirable event (local conflict continuation in the example), the probability of the policy realizing at least this level of reduction, and the ratio of the cost of the policy to the cost of the undesirable event.

\section{Conclusion}
This article has introduced the causal binary loss function model, a general Bayesian decision-theoretic model for binary outcomes, to evaluate the practical implications of the effect of a policy on an undesirable event. The model indicates that whether a causal factor should be implemented as a policy to reduce the likelihood of an undesirable event, depends on the effect size range of interest, its probability mass, the cost of the policy, and the cost of the event. It provides richer information about the practical implications of the statistically estimated causal effects of policy interventions, than the significance test using only a standard point estimate and (categorical) uncertainty measures such as the dichotomy of statistical significance vs. insignificance. 

While the causal binary loss function model focuses on research on policy interventions, it also has implications for non-policy research. It might be argued that non-policy research should use criteria common in the academic community, such as statistical significance and Cohen's effect size measure \autocite{Cohen1988}, to detect an academically meaningful effect, and should be agnostic about a utility/loss function for policymaking. There are two issues to be considered, however. First, it is well-known that the use of statistical significance has led to a skewed distribution of knowledge -- the publication bias \autocite{Esarey2016, Gerber2008, Simonsohn2014}. Second, there is no consensus over what an academically meaningful effect size is. It is unclear how one can define a meaningful effect size without reference to practical contexts. One solution to these two issues is to use a general decision-theoretic model, such as mine, which treats the continuous nature of statistical uncertainty as such and evaluates effect sizes based on a loss function.

The article has several implications for future research. First, if more than one statistical model is conceivable (which is usually the case), there is uncertainty over the crossover point where an expected loss is smaller if the policy is implemented than if it is not. Such uncertainty could be mitigated by incorporating into the loss function the results aggregated from all conceivable statistical models \autocite{Montgomery2010}.

Second, the causal binary loss function may be extended to be used for (ordered or non-ordered) categorical outcome variables. This could be done by computing the expected loss given the cost of each category of the outcome (e.g., a peace agreement, a ceasefire agreement, a stalemate, and ongoing conflict) and the effect of a policy intervention on each outcome category.

Third, the causal binary loss function may also be extended to analyze the effect of more than one policy on an undesirable event. This is simplest when multiple policies are causally independent from one another in the data used. We use the variables to capture these policies in one logistic regression, compute $p\theta_{int}$ and $p\theta_{unint}$ for each policy, and plug all of them into the loss function. The cost of policy implementation should be that of implementing all these policies. Meanwhile, if some of the multiple policies cause, or are caused by, each other in the data used, more careful statistical causal modeling \autocite[e.g., mediation analysis; see][]{Imai2010} will be necessary to estimate the independent effect of each policy before using their posteriors in the loss function.

Fourth, a policy may influence more than one outcome. In the example used here, the model considers only the (average) effect of peacekeepers on the likelihood of local conflict continuation. But peacekeeping, and many other policies, can affect multiple aspects of society. There can be more benefits of peacekeeper deployment than a reduction in the likelihood of local conflict continuation, e.g., greater educational attainment \autocite{Reeder2021} or better environmental quality \autocite{Bakaki2021}. Peacekeeper deployment might also have an adverse effect on some other aspects, e.g., sexual violence by peacekeepers \autocite{Nordas2013}. To incorporate the effect of a policy intervention on multiple outcomes, a statistical model and the causal binary loss function could be applied to each outcome, and then the expected losses could be aggregated over all models.

Fifth, a framework for multi-phase decision analysis could be developed. If a policy were implemented, a follow-up analysis could use the actual consequence of the policy as new data and the last posterior estimate as a new prior, thereby updating the posterior and therefore the estimate of the expected loss.

Finally, should the causal binary loss function model serve actual policymaking, for example, to compare the relative cost-effectiveness of several policy options under a certain budget constraint, the question would be how to measure costs, which are usually unobservable and difficult to quantify. A Bayesian statistical model could be developed to estimate costs as latent variables, and their posterior predicted values could be plugged into the loss function.

\section*{Acknowledgments}
I am grateful for their helpful comments to the editors and the anonymous reviewers, Johan A. Elkink, Nial Friel, Zbigniew Truchlewski, Martijn Schoonvelde, James Cross, and participants in the 2019 GPSA-SPSA Methodology conference, the 2019 Bayesian Nonparametrics conference, the 2019 Max Weber June Conference, the 2019 PSAI Annual Conference, and in seminars at Dublin City University, University College Dublin, and the University of Strathclyde. I would like to acknowledge the receipt of funding from the Irish Research Council (the grant number: GOIPD/2018/328) for the development of this work. The views expressed are my own unless otherwise stated, and do not necessarily represent those of the institutes/organizations to which I am/have been related.

\section*{Supplemental Materials}
The supplemental document, and the R code to reproduce the results in this article, are available at \url{https://akisatosuzuki.github.io/papers.html}. A step-by-step vignette to install the R package ``bayesdtm'' and implement the causal binary loss function model is available at \url{https://akisatosuzuki.github.io/bayesdtm.html}.

\section*{Declaration of interest statement}
The author declares no conflict of interest.

\printbibliography

\end{document}